\documentclass[twocolumn]{aastex62}

\pdfoutput=1
\usepackage{natbib}

\submitjournal{ApJL}

\shorttitle{Indication of Another IMBH in the Galactic Center}
\shortauthors{Takekawa et al.}

\begin{document}

\title{Indication of Another Intermediate-mass Black Hole in the Galactic Center}

\correspondingauthor{Shunya Takekawa}
\email{shunya.takekawa@nao.ac.jp}

\author{Shunya Takekawa}
\affil{Nobeyama Radio Observatory, National Astronomical Observatory of Japan\\
462-2 Nobeyama, Minamimaki, Minamisaku-gun, Nagano 384-1305, Japan}

\author{Tomoharu Oka}
\affiliation{Department of Physics, Institute of Science and Technology, Keio University\\
3-14-1 Hiyoshi, Kohoku-ku, Yokohama, Kanagawa 223-8522, Japan}
\affiliation{School of Fundamental Science and Technology, Graduate School of Science and Technology, Keio University\\
3-14-1 Hiyoshi, Kohoku-ku, Yokohama, Kanagawa 223-8522, Japan}

\author{Yuhei Iwata}
\affiliation{School of Fundamental Science and Technology, Graduate School of Science and Technology, Keio University\\
3-14-1 Hiyoshi, Kohoku-ku, Yokohama, Kanagawa 223-8522, Japan}

\author{Shiho Tsujimoto}
\affiliation{School of Fundamental Science and Technology, Graduate School of Science and Technology, Keio University\\
3-14-1 Hiyoshi, Kohoku-ku, Yokohama, Kanagawa 223-8522, Japan}

\author{Mariko Nomura}
\affiliation{Astronomical Institute, Tohoku University\\
6-3 Aramaki, Aoba, Sendai 980-8578, Japan}

\begin{abstract}
We report the discovery of molecular gas streams orbiting around an invisible massive object in the central region of our Galaxy, based on the high-resolution molecular line observations with the Atacama Large Millimeter/submillimeter Array (ALMA).
The morphology and kinematics of these streams can be reproduced well through two Keplerian orbits around a single point mass of $(3.2\pm0.6)\times 10^4$ $M_\odot$.
We also found ionized gas toward the inner part of the orbiting gas, indicating dissociative shock and/or photoionization. 
Our results provide new circumstantial evidences for a wandering intermediate-mass black hole in the Galactic center, suggesting also that high-velocity compact clouds can be probes of quiescent black holes abound in our Galaxy.
\end{abstract}


\section{Introduction}
Intermediate-mass black holes (IMBHs) with masses of $10^2$--$10^5$ $M_\odot$ are the missing link between stellar-mass and supermassive black holes \citep[e.g.,][]{ebisuzaki01}.
Many efforts have been made to confirm the existence of IMBHs \citep{mezcua17}.
Ultra-luminous X-ray sources have been considered as promising IMBH candidates \citep[e.g.,][]{farrell09}.
Relatively massive IMBHs may lurk in the nuclei of dwarf galaxies and/or globular clusters \citep[e.g.,][]{reines13,baldassare15,kiziltan17}. 
However, these results have been argued upon, and therefore, none of the IMBH candidates are accepted as definitive \citep[e.g.,][]{ebisawa03,strader12}.

High-velocity(-width) compact clouds (HVCCs), which are a population of compact molecular clouds with extremely broad velocity widths \citep{oka98,oka99}, may also provide possible hints of the existence of IMBHs. 
HVCC CO--0.40--0.22 has been interpreted as a cloud that was gravitationally kicked by a massive IMBH with a mass of $10^5$ $M_\odot$ \citep{oka16,oka17}.
Although the putative IMBH has a radio emission counterpart CO--0.40--0.22$^*$, the nature of CO--0.40--0.22 and CO--0.40--0.22$^*$ is still controversial \citep{ravi18,tanaka18}.
IRS13E, which is an infrared source in the vicinity of the Galactic nucleus Sgr A$^*$, has also been suggested to contain a $\sim 10^4$ $M_\odot$ IMBH based on the high-velocity feature of the ionized gas \citep{tsuboi17} as well as the stellar dynamics \citep{schodel05}.

Recently, we discovered a peculiar HVCC, HCN--0.009--0.044, at a projected distance of 7 pc from Sgr~A$^*$, by using the James Clerk Maxwell Telescope (JCMT) \citep{takekawa17}. 
HCN--0.009--0.044 is more compact ($\sim 1$ pc) than any previously known HVCCs (2--5 pc), and its velocity width ($\sim40$ km s$^{-1}$) is typical to those of HVCCs.
The compactness, kinematics, and absence of luminous stellar counterpart can be explained by the high-velocity plunge of an invisible compact object into a molecular cloud \citep{takekawa17,nomura18}.
The driving source may be an inactive and isolated black hole.

This Letter reports on the results of high-resolution observations of HCN--0.009--0.044 with the Atacama Large Millimeter/submillimeter Array (ALMA) and the discovery of molecular gas streams showing clear orbital motions around an invisible gravitational source with a mass of $(3.2\pm0.6)\times 10^4$ $M_\odot$. 
The distance to the Galactic center is assumed to be $D=8$ kpc.

\section{Observations}
Our ALMA cycle 5 observations (2017.1.01557.S) were performed on May 13--14, 2018.
Eleven 7-m antennas and forty-six 12-m antennas were used to obtain the HCN {\it J}=4--3 (354.5 GHz), HCO$^+$ {\it J}=4--3 (353.6 GHz), and CS {\it J}=7--6 (342.9 GHz) datasets of the target source HCN--0.009--0.044.
{The field of view was $54\arcsec \times 54\arcsec$ centered at $(l,\ b)=(-0.009\arcdeg,\ -0.044\arcdeg)$, which was covered with 7 and 23 pointings of the 7-m and 12-m arrays, respectively.
The on-source times were 43.34 and 9.27 min for the 7-m and 12-m array observations, respectively.
The 12-m array configuration was C43-2 with baseline lengths of 15--314 m.
The bandwidths of the spectral windows for the HCN, HCO$^+$, and CS observations were respectively 1, 0.5, and 2 GHz with 1.953-MHz channel widths.
}
J1924--2914 and J1517--2422 were observed as the flux and bandpass calibrators.
The phase calibrators were J1744--3116 and J1700--2610.

{We calibrated and reduced the data by using the Common Astronomy Software Applications (CASA) software package (version 5.1.2-4) in the standard manner\footnote{\url{https://casaguides.nrao.edu/index.php/ALMAguides}}.
The calibration was performed with the calibration script provided by the East Asian ALMA Regional Center.}
The visibility data were split into spectral lines and continuum emissions through the task `uvcontsub'.
The synthesized images were created using the task `tclean' with Briggs weighting {with a robust parameter of 0.5}.
The spatial and velocity grid width of the resultant image cubes were 0.5$''$ and 2 km s$^{-1}$, respectively.
The synthesized beam sizes were as follows: $0.87''\times 0.71''$ with a position angle (PA) of $-31.6^\circ$ at 354.5 GHz, $0.87''\times 0.71''$ with a PA of $-30.3^\circ$ at 353.6 GHz, and $0.89''\times 0.73''$ with a PA of $-30.7^\circ$ at 342.9 GHz.
The root mean square (rms) noise level of the image cubes was 7 mJy beam$^{-1}$.

\section{Results and Discussion}
\subsection{Spatial/Velocity Structure and Physical Condition}
Our high-resolution observations have unveiled the entity of HCN--0.009--0.044.
Figure 1(a) shows the integrated intensity map of the HCN {\it J}=4--3 line.
HCN--0.009--0.044 is resolved into several components.
The main body appears as a balloon-like structure (hereafter Balloon).
A stream-like structure (hereafter Stream) lies on the southeast of the Balloon.
An ultra-compact clump (UCC) is associated with the tip of the Stream.
Figure 1(b) shows the averaged line-of-sight velocity (moment 1) map of the HCN line.
The Balloon and Stream clearly exhibit velocity gradients.
The gas of the Balloon is gradually accelerated clockwise from the northeastern part at the line-of-sight velocity of $V_{\rm LSR} \sim -70$ km s$^{-1}$, indicating a rotational motion (see also Figure 2).
The Stream is accelerated from south to north ( $V_{\rm LSR} \sim -60$ km s$^{-1}$ to $-40 $ km s$^{-1}$) as if it is orbiting around the center of the Balloon.

\begin{figure*}
\begin{center}
\includegraphics[width=16cm]{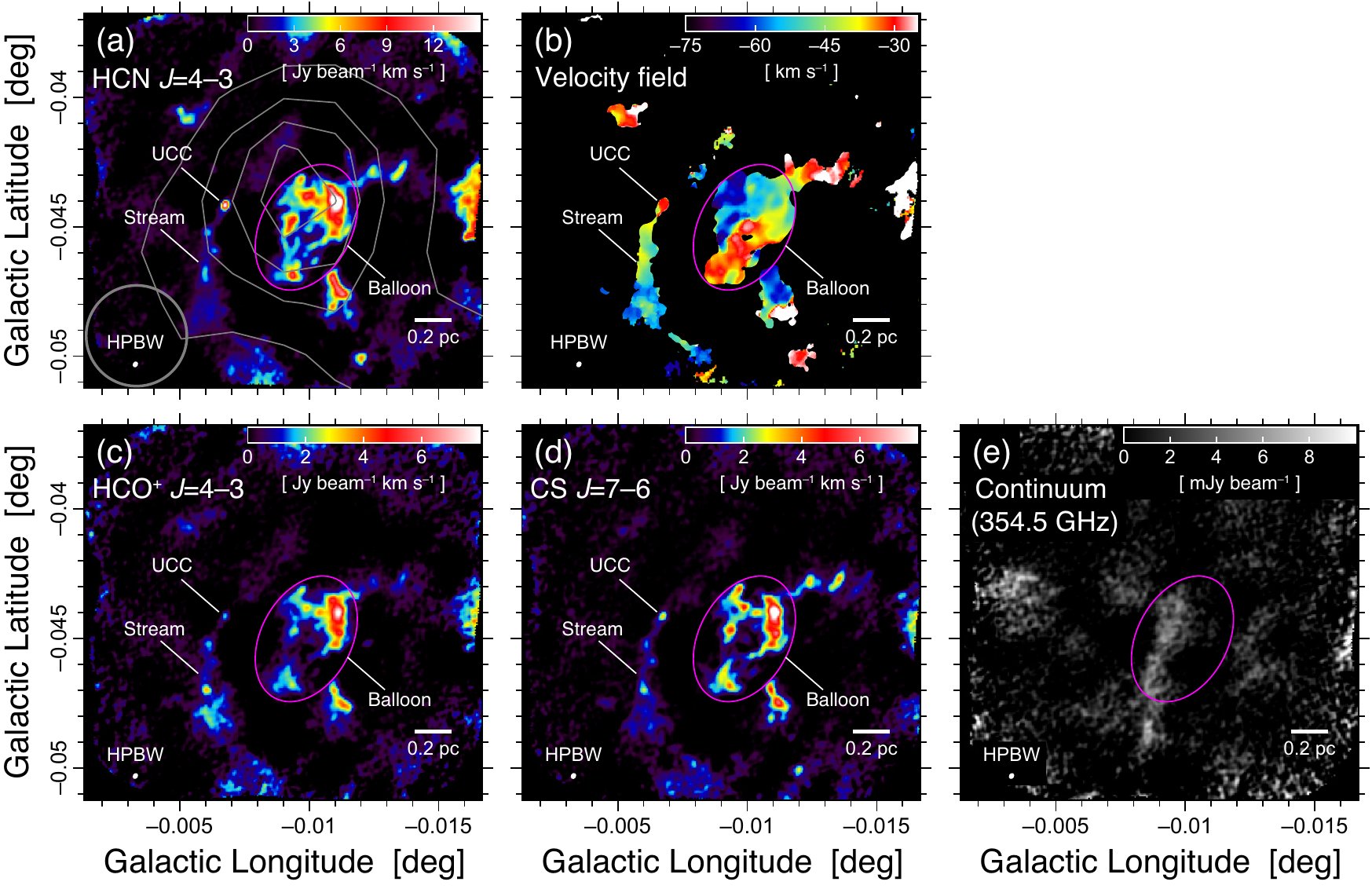}
\caption{
(a) Integrated intensity map of the HCN {\it J}=4--3 line over $V_{\rm LSR}=-80$ -- $-20\ \rm km\ s^{-1}$.
The gray contours show the same map obtained by using the JCMT \citep{takekawa17}.
The magenta ellipse indicates the extent of the Balloon.
The half-power beam widths (HPBWs) of ALMA and JCMT are represented by a white filled ellipse and gray circle, respectively.
(b) Averaged line-of-sight velocity (moment 1) map of the HCN line.
{
{(c, d) Integrated intensity maps of the HCO$^+$ {\it J}=4--3 and CS {\it J}=7--6 lines over $V_{\rm LSR}=-80$ -- $-20\ \rm km\ s^{-1}$.}
(e) Continuum image at 354.5 GHz obtained with the 1-GHz bandwidth.
{The rms noise level of the continuum is 2 mJy beam$^{-1}$.}}
}
\end{center}
\end{figure*}

\begin{figure*}
\begin{center}
\includegraphics[width=16cm]{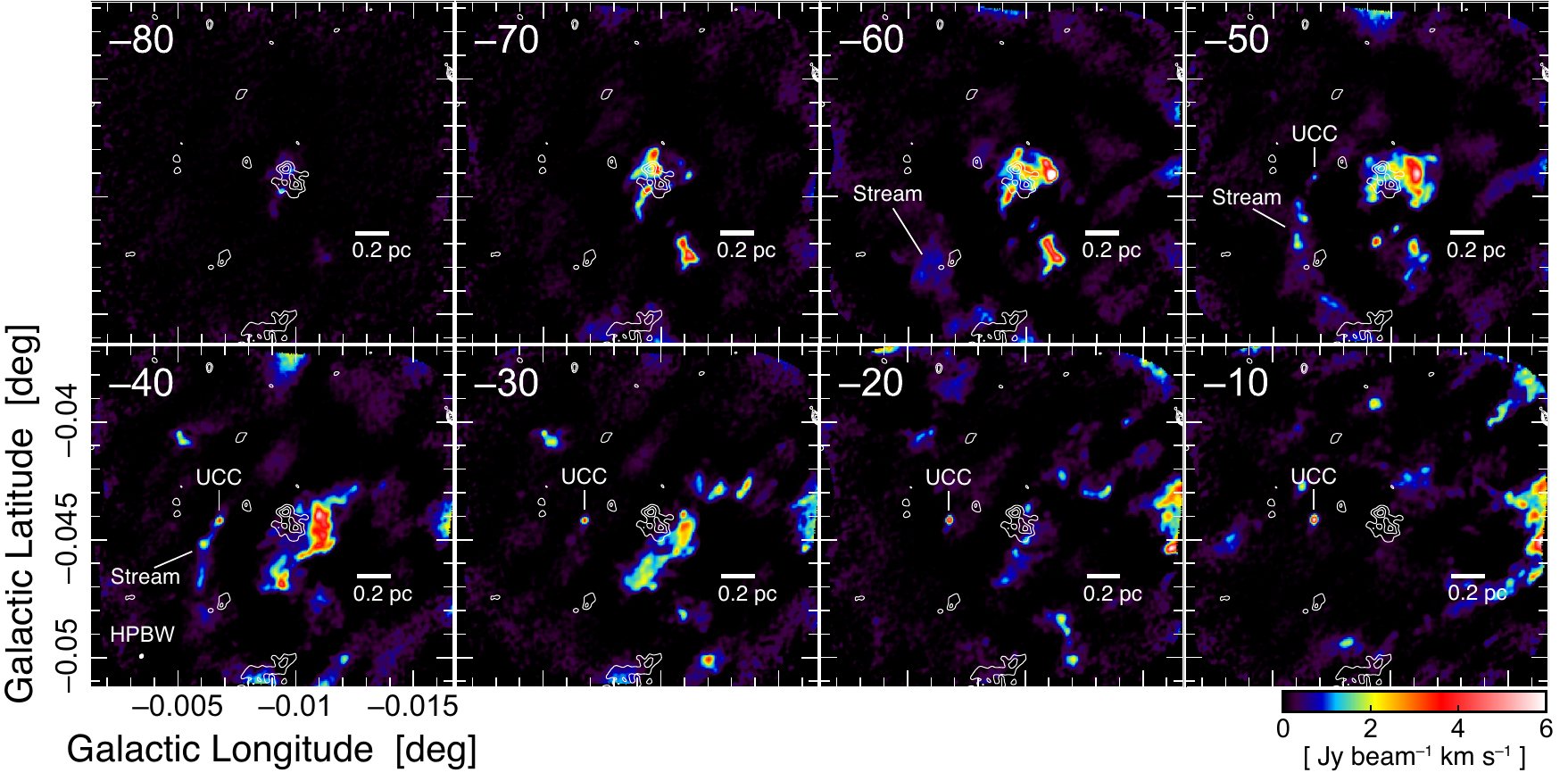}
\caption{Velocity channel maps of HCN--0.009--0.044.
Each panel shows the HCN {\it J}=4--3 emission integrated over $\rm 10\ km\ s^{-1}$.
The numerical value on the upper left corner in each panel indicates the central velocity in $\rm km\ s^{-1}$.
The white filled ellipse on the lower left corner indicates the HPBW of ALMA.
The white contours show the P$\alpha$ emission at 1.87 $\mu\rm m$ \citep{dong11}. 
The contour levels are 300, 350, and 400 $\mu \rm Jy\ arcsec^{-2}$.
The rms noise level of the P$\alpha$ map is $\sim 74\ \mu \rm Jy\ arcsec^{-2}$.
}
\end{center}
\end{figure*}

HCN--0.009--0.044 is likely to be at least 5 kpc apart from the Sun ($R > 5$ kpc) because CO {\it J}=3--2 emission from HCN--0.009--0.044 \citep{parsons18} suffers from absorption by molecular gas in the 3-kpc arm, which is the Galactic spiral arm at $\sim 5$ kpc \citep{sofue06}.
The components traced by the HCN {\it J}=4--3 emission also appear in the ALMA images of the HCO$^+$ {\it J}=4--3 and CS {\it J}=7--6 lines {(Figures 1(c) and (d))}.  
HCN--0.009--0.044 consists of highly dense and hot molecular gases as the critical densities for these transitions are $\sim 10^7$ cm$^{-3}$, and the {\it J}=7 level of CS is 65.8 K above the ground state. 
Such dense/hot molecular gases are abundant in the central 200 pc of our Galaxy but rare in the Galactic disk region.
Hence, HCN--0.009--0.044 is probably in the Galactic center ($R=8$ kpc).
The molecular gas masses of the Balloon, Stream {and UCC} were estimated to be $M_{\rm LTE} \sim 6$ $M_\odot$, 1 $M_\odot$ {and 0.1 $M_\odot$}, respectively, assuming the local thermodynamic equilibrium (LTE) with an excitation temperature of 22 K in the optically thin limit \citep{takekawa17} and a fractional abundance of $\rm [HCN]/[H_2] = 4.8\times 10^{-8}$ \citep{tanaka09,oka11}.

{Figure 1(e) shows the continuum image at 354.5 GHz.
A faint filamentary structure appears in the eastern side of the Balloon.
Although this filament also appears in the 353.6- and 342.9-GHz bands, we could not derive the reliable spectral index because of the insufficient sensitivities.
M--0.02--0.07 (also referred to as the +50 km s$^{-1}$ cloud) and the Northern Ridge at $V_{\rm LSR}\simeq 0$ km s$^{-1}$ \citep{mcgary01} overlap with the filament in the line of sight.
The filament may reflect dust emission from them. 
The physical relation between the continuum filament and the HVCC is currently ambiguous.
}

\subsection{Origin of HCN--0.009--0.044}
Interactions with supernova remnants \citep{oka08,tanaka09,yalinewich18} and cloud--cloud collisions \citep{tanaka15,tanaka18,ravi18} have been considered as the origins of HVCCs.
However, the position--velocity structure of HCN--0.009--0.044 exhibits neither expanding motion indicative of a supernova--cloud interaction {\citep[e.g., see Figure 12 in ][]{fukui12}} nor V-shaped pattern characteristic of a cloud--cloud collision {\citep[e.g., see Figure 10(c) in ][]{torii17}}.
Natural explanations for the velocity gradients along the Balloon and Stream (Figure 1(b)) are orbital motions caused by gravity.
The observed kinematics implies that a massive gravitational source lurks in the Balloon.
{The detection of the CS {\it J}=7--6 line (Figure 1(d)), which is a high-density and possibly shock probe \citep[e.g.,][]{tanaka_18}, may support the interpretation that the molecular gas has experienced the tidal compression by the strong gravity.}
The enclosed mass ($M$) can be roughly estimated by $M \sim {\Delta V}^2 R {\Delta \theta}/2G$, where $\Delta V$ is the velocity width, $R$ is the distance to the cloud, $\Delta \theta$ is the angular diameter, and $G$ is the gravitational constant.
The observed velocity width and diameter of the Balloon are $\Delta V \sim 20$ km s$^{-1}$ and $R\Delta \theta \sim 0.4$ pc, respectively.
Thus, a massive object with $\sim 10^4$ $M_\odot$ may be hidden in the Balloon.

\begin{table*}
 \caption{Parameters of the modeled Keplerian orbits}
 \centering
  \begin{tabular}{lll}
   \hline \hline 
   Parameters &  Balloon & Stream \\
   \hline
   Semi-major axis, $a$ &  $0.20 \pm 0.02$ pc &  $0.54 \pm 0.15$ pc \\
   Eccentricity, $e$ & $0.66 \pm 0.05$ & $0.62 \pm 0.31$\\
   Longitude of ascending node, $\Omega$ & $62 \pm 24$ deg  &  $-113 \pm 41$ deg  \\
   Argument of pericenter, $\omega$ & $-99 \pm 21$ deg &  $-115 \pm 34$ deg \\
   Inclination\tablenotemark{\rm a}, $i$ &  $153 \rm \ (or\ 27) \pm 13$ deg & $136 \rm \ (or\ 44) \pm 21$ deg  \\
   Pericenter distance &   $0.07$ pc &   $0.21$ pc \\
   Orbital period &  $4.6\times10^4$ years &  $21.1\times10^4$ years \\
   \hline
  \end{tabular}
  \tablenotetext{\rm a}{ If the orbital motion is counterclockwise, the inclination of $i < 90\arcdeg$ is chosen.}
\end{table*}

\subsection{{Keplerian Model}}
{In order to confirm whether Keplerian motions explain the kinematical structures of the Balloon and Stream, we performed orbital fittings to the observed data with the similar procedure as done for the Sgr A West by Zhao et al. (2009).
Assuming the dynamical center at $(l,\ b, \ R) = (-0.0096^\circ,\ -0.0451^\circ,\ \rm 8\ kpc)$, we determined three-dimensional orbital parameters ($a$, $e$, $\Omega$, $\omega$, $i$) for the orbital geometries of the Balloon and Stream by least-square fitting to the loci we chose (red and blue points in Figure 3(b)).
{We carefully chose these loci (including the dynamical center) so that they represent the kinematics of the observed features.
}
The orbital parameters are the semi-major axis ($a$), eccentricity ($e$),  longitude of ascending node ($\Omega$), argument of pericenter ($\omega$), and inclination angle ($i$).
We set the {\it X}-,  {\it Y}-, and {\it Z}-axes parallel to the Galactic longitude, Galactic latitude, and line-of-sight direction, respectively.
The best-fit orbital parameters are listed in Table 1 and the resultant orbits are projected in Figures 3(a)--(c).
{Note that there remains a double degeneracy in the inclination angle.
Orbits with $i$ and $(180\arcdeg - i)$ produce the same projected orbits and line-of-sight velocities.
}

The line-of-sight velocity ($V_Z$) at each ({\it X}, {\it Y}) of the orbits depend on the mass ($M_{\rm dyn}$) and line-of-sight velocity ($V_{\rm dyn}$) of the dynamical center.
After determining the orbital geometries, we fitted the modeled velocities $V_Z$ to the observed velocities on the orbits of the Balloon and Stream with free parameters of $M_{\rm dyn}$ and $V_{\rm dyn}$ through a chi-square ($\chi^2$) minimization approach.
We used the moment-1 values of the HCN image smoothed with a Gaussian kernel of 3$''$ full width at half maximum (Figure 3(a)) for the model fit to reduce the noise effect.
As a result, we derived the best-fit values of $M_{\rm dyn} = (3.2\pm0.6) \times 10^4$ $M_\odot$ and  $V_{\rm dyn} = -49.5^{+1.0}_{-0.7} $ km s$^{-1}$.
Figure 3(c) shows the velocity field of the model with the best-fit parameters.
Figure 3(d) shows the confidence level contours as a function of  $M_{\rm dyn}$ and  $V_{\rm dyn}$.
These suggest that the observed kinematics of the Balloon and Stream can be well explained by two Keplerian orbits around a single gravitational source with $M_{\rm dyn} \simeq 3\times 10^4$ $M_\odot$.
}

\begin{figure*}
\begin{center}
\includegraphics[width=16 cm]{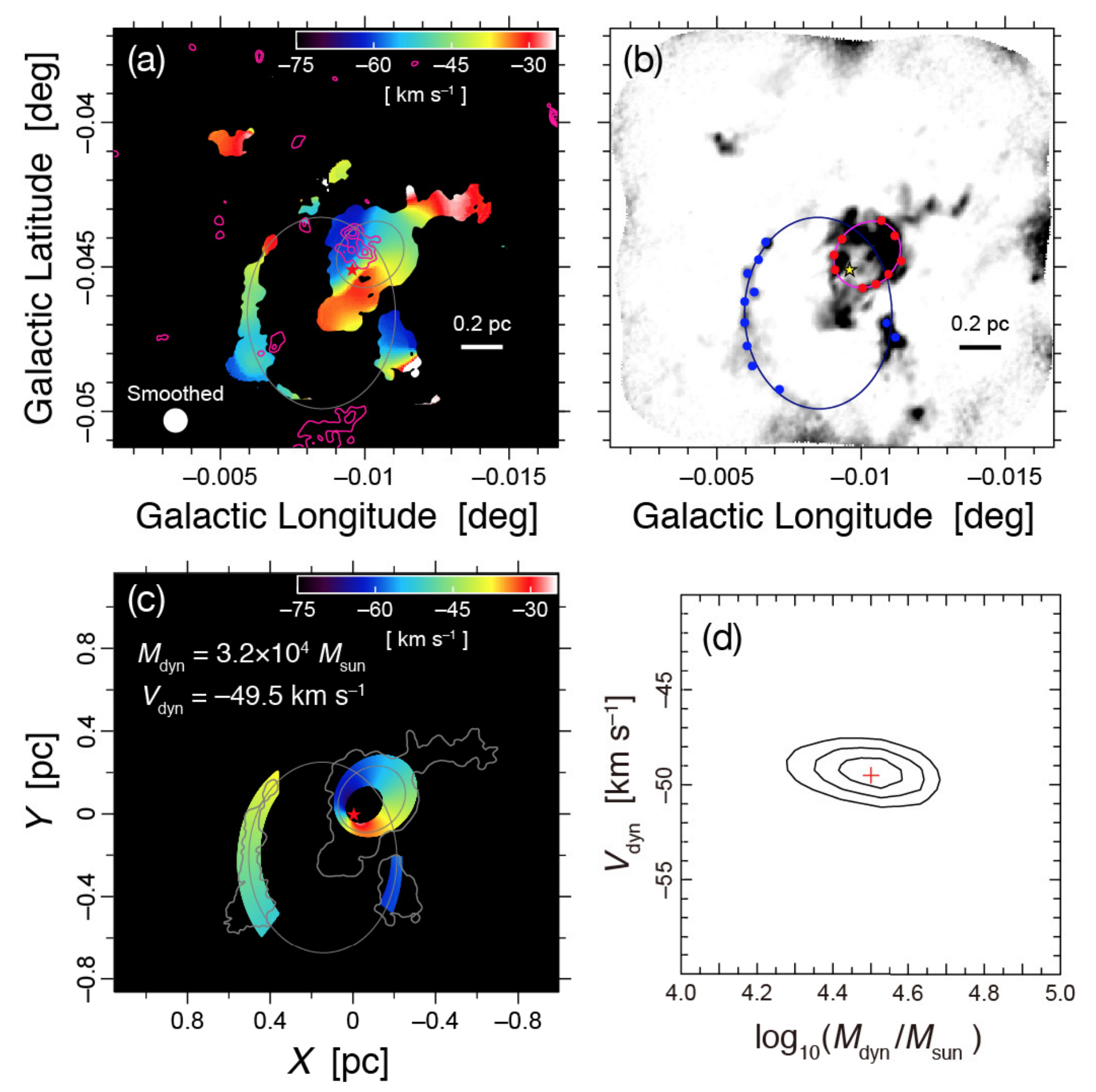}
\caption{
(a) Averaged line-of-sight velocity (moment 1) map of the HCN line smoothed with a 3$''$ Gaussian kernel.
The white filled circle on the lower left corner indicates the full width at half maximum of the Gaussian kernel.
The magenta contours show the P$\alpha$ emission \citep{dong11}. 
{
(b) Loci used for the orbital fittings for the Balloon (red) and Stream (blue).
The background image (grayscale) is the HCN integrated intensity map.
The ellipses are the modeled orbits for the Balloon and Stream, respectively.
}
Their orbit parameters are listed in Table 1.
The star indicates the dynamical center.
(c) Color map of the line-of-sight velocities ($V_Z$) of the modeled orbits with the best-fit values.
The contours show the HCN intensity of 1 $\rm Jy\ beam^{-1}\ km\ s^{-1}$.
{
The ellipses and star are the same as those in the panel (b).
(d)  Confidence level contours as a function of $M_{\rm dyn}$ and  $V_{\rm dyn}$.
The contour levels are $1\sigma$ (68.3\%), $2\sigma$ (95.4\%), and $3\sigma$ (99.7\%).
The cross mark indicates the best-fit values.
}
}
\end{center}
\end{figure*}

{
 Figure 4 shows the three-dimensional  schematic view of HCN--0.009--0.044 based on the model with the best-fit parameters.}
The cloud lying on the south of the Balloon with $V_{\rm LSR}\sim -60$ km s$^{-1}$ may share the same orbit as the Stream.
The dynamical time scales of the Balloon and Stream are estimated to be $4\times 10^4$ and $6\times 10^4$ years, respectively.
The similarity of the dynamical time scales may indicate that the Balloon and Stream are derived from a common parent molecular cloud.
These results are consistent with the notion that the Balloon and Stream have been captured by the gravitational potential well of a massive compact object in the relatively recent past.

{It should be noted that the UCC shows an abnormally broad velocity width which can not be explained by the Keplerian model.
The UCC is located at the northern tip of the Stream, having velocities from $V_{\rm LSR}\simeq -50$ km s$^{-1}$ to 0 km s$^{-1}$ without velocity gradient (Figure 2).
The positive velocity end is contaminated with emission from the Northern Ridge \citep{mcgary01}.
The virial mass is roughly estimated to be $\sim 10^4$ $M_\odot$, which is far greater than the LTE mass ($\sim 0.1$ $M_\odot$).
The UCC may contain a stellar object as a core, and the broad velocity width could be attributed to an outflow from it.
Further observations with higher spatial resolution are necessary to reveal the origin of the UCC.
}

\begin{figure}
\begin{center}
\includegraphics[width=8.5 cm]{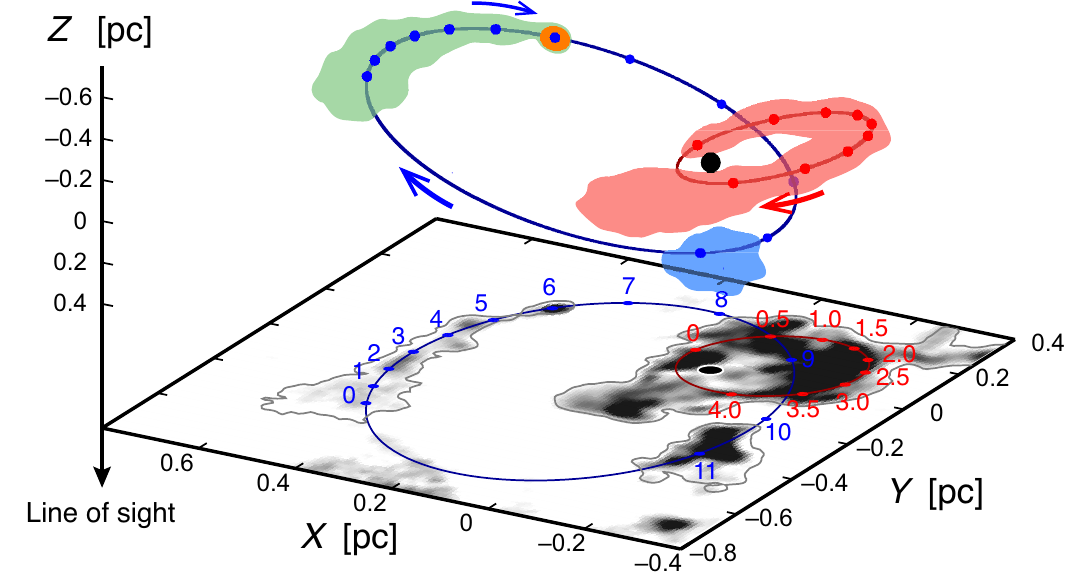}
\caption{
 {
 Three-dimensional schematic view of HCN--0.009--0.044 based on the Keplerian model with the best-fit parameters.
 We adopt clockwise motions ($i > 90\arcdeg$) for the orbits.
 The red and blue arrows indicate the direction of the orbital motions for the Balloon and Stream, respectively,
 The red, green, blue and orange clouds represent the Balloon, Stream, $-60$ km s$^{-1}$ clump, and UCC, respectively.
 The black circle indicates the position of the IMBH.
 The red and blue points on the orbits, which are also projected on the {\it X}--{\it Y} plane, indicate the elapsed times.
 The red and blue numerical values are the elapsed times in unit of $10^4$ yr.
 The grayscale on the {\it X}--{\it Y} plane is the HCN map.
}
}
\end{center}
\end{figure}

\subsection{Indication of an IMBH}
According to the model, a huge mass of $(3.2\pm0.6) \times 10^4$ $M_\odot$ must be concentrated within a radius significantly smaller than 0.07 pc (the pericenter distance for the Balloon's orbit). 
The averaged mass density is much larger than that of the core of M15 ($\rho \sim 2\times 10^7$ $M_\odot$ pc$^{-3}$), which is one of the most densely packed globular clusters \citep{djorgovski84}.
However, considering the star cluster as the gravitational source is implausible because of the absence of luminous stellar counterparts.
Therefore, the most promising candidate for the gravitational source is a massive IMBH.
Although accreting black holes can be traced through X-ray emission from their accretion disks \citep[e.g.,][]{done10}, no X-ray point source has been detected in the extent of the Balloon \citep{muno09}.
This is possibly because the IMBH is isolated and inactive. 
HCN--0.009--0.044 is the third case of the possible IMBH holder in the Galactic center after IRS13E \citep{schodel05,tsuboi17} and CO--0.40--0.22 \citep{oka16,oka17}.

{Many black holes follow the Fundamental Plane of black hole activity, which is an empirical relation between radio luminosity, X-ray luminosity and black hole mass \citep[e.g.,][]{merloni03, falcke04}}.
{Diffuse continuum emission at 5.5 GHz \citep{zhao13,zhao16} overlaps with HCN--0.009--0.044 \citep[see Figure 2(b) in][]{takekawa17}.
This sensitive image provides an upper limit radio luminosity of $\sim 2\times 10^{28}$ erg s$^{-1}$ for the putative IMBH.
If the IMBH is placed on the Fundamental Plane derived by Plotkin et al. (2012), then the X-ray luminosity can be inferred to be $\lesssim 10^{31}$ erg s$^{-1}$ from the radio luminosity of $\lesssim 2\times 10^{28}$ erg s$^{-1}$ and the mass of $3.2\times10^4$ $M_\odot$.
On the other hand, the X-ray image at 0.5--8 keV \citep{muno09} provides an upper limit X-ray luminosity of $\sim 7\times 10^{30}$ erg s$^{-1}$ for the IMBH.
This is consistent with the Fundamental Plane for low accretion rate black holes \citep{plotkin12}, {thereby providing another support for the IMBH interpretation.}
}

Although there is no point source suggestive of the IMBH, we found an emission counterpart for the Balloon in the P$\alpha$ recombination line at 1.87 $\mu$m obtained through the Hubble Space Telescope \citep{dong11} (Figures 2 and 3(a)).
The P$\alpha$ emitting region seems to be surrounded by the orbiting molecular gas.
This probably suggests that the inner part of the Balloon's orbit has been ionized.
The ionization may be attributed to the dissociative shock caused by the sudden acceleration of molecular gas by the strong gravitational force.
Hence, the detection of the P$\alpha$ emission may also support for the presence of the IMBH{, although a stellar emission as the origin of the P$\alpha$ cannot be ruled out.
}

{
Several observational studies have reported interactions of outflows and/or intense radiation from non-nuclear black holes with their ambient gas \citep[e.g.][]{soria14, tetarenko18}.
For instance, a persistent jet from Cygnus X-1 has been suggested to create a shell-like structure of ionized gas \citep{gallo05, russell07}.
In addition, the elongated ionized gas feature associated with M51 ULX-1 has been considered as evidence for an interaction with a black hole jet \citep{urquhart18}.
The P$\alpha$ emission feature we noticed could indicate a past activity of the IMBH although its spatial distribution shows neither shell-like nor elongated morphology.
}
Further investigations of the kinematics {and physical conditions} of the ionized gas would provide more convincing evidence for the presence of an IMBH.

Several globular clusters and dwarf galaxies have been suggested to harbor massive black holes with masses lesser than $10^6$ $M_\odot$ as their nuclei \citep{reines13,baldassare15,kiziltan17}. 
Recently, it has been indicated that there is likely to be a $4\times 10^4$ $M_\odot$ IMBH in the center of $\omega $ Centauri, which is the most luminous globular cluster in our Galaxy \citep{baumgardt18}.
The IMBH in HCN--0.009--0.044 could be a remnant of such a globular cluster.
The parent cluster have possibly already been dissolved before falling into the Galactic center.

The detected rotational motion of HCN--0.009--0.044 strongly indicates the presence of a dark massive object, which is probably an IMBH.
An emission counterpart of the IMBH could be identified through future multi-wavelength observations.
Similar to the simulations conducted for HVCC CO--0.40--0.22 \citep{ballone18}, the hydrodynamical simulations of the molecular clouds would more accurately restrict the orbital geometries and IMBH mass.
Although only $\sim 60$ black holes have been identified in our Galaxy to date \citep{corral-santana16}, the total number of black holes in our Galaxy is theoretically estimated as $\sim 10^8$--$10^9$ \citep{agol02,caputo17}.
{Additionally, at least $10^4$ black hole X-ray binaries are observationally inferred to lurk in our Galaxy \citep{tetarenko16}.}
These suggest that almost all black holes are inactive with low accretion rates.
Regardless of the activity status of a black hole, its strong gravity can disturb the ambient gas.
As shown, high-resolution observations of compact high-velocity gas features have the potential to increase the number of candidates for non-luminous black holes, providing a new perspective to search for the missing black holes.

\acknowledgments
This paper makes use of the following ALMA data: ADS/JAO.ALMA\#2017.1.01557.S.
 ALMA is a partnership of ESO (representing its member states), NSF (USA) and NINS (Japan), together with NRC (Canada), MOST and ASIAA (Taiwan), and KASI (Republic of Korea), in cooperation with the Republic of Chile. The Joint ALMA Observatory is operated by ESO, AUI/NRAO and NAOJ.
We are grateful to the ALMA staff for conducting the observations and providing qualified data.
We are also thankful to the anonymous referee for helpful comments and suggestions that improved this paper.
\facility{ALMA}
\software{CASA (version 5.1.2)}


\begin{thebibliography}{1}
\bibitem[Agol \& Kamionkowski 2002]{agol02} Agol, E. \& Kamionkowski, M. 2002, \mnras, 334, 553
\bibitem[Baldassare et al. 2015]{baldassare15} Baldassare, V. F., Reines, A. E., Gallo, E. \& Greene, J. E. 2015, \apjl, 809, L14
\bibitem[Ballone et al. 2018]{ballone18} Ballone, A., Mapelli, M., \& Pasquato, M. 2018, \mnras, 480, 4684
\bibitem[Baumgardt \& Hilker 2018]{baumgardt18} Baumgardt, H. \& Hilker, M. 2018, \mnras, 478, 1520
\bibitem[Caputo et al. 2017]{caputo17} {Caputo}, D.~P., {de Vries}, N., {Patruno}, A. \& {Portegies Zwart}, S. 2017, \mnras, 468, 4000
\bibitem[Corral-Santana et al. 2016]{corral-santana16} Corral-Santana, J. M., Casares, J., Mu{\~n}oz-Darias, T., et al. 2016, \aap, 587, 61
\bibitem[Djorgovski \& King 1984]{djorgovski84} Djorgovski, S. \& King, I. R. 1984, \apjl, 277, L49
\bibitem[Done 2010]{done10} Done, C. 2010, arXiv:1008.2287
\bibitem[Dong et al. 2011]{dong11} Dong, H., Wang, Q. D., Cotera, A., et al. 2011, \mnras, 417, 114
\bibitem[Ebisawa et al. 2003]{ebisawa03} {Ebisawa}, K., {{\.Z}ycki}, P., {Kubota}, A., {Mizuno}, T. \& {Watarai}, K.-y. 2003, \apj, 597, 780
\bibitem[Ebisuzaki et al. 2001]{ebisuzaki01} Ebisuzaki, T., Makino, J., Tsuru, T. G., et al. 2001, \apjl, 562, L19
\bibitem[Falcke et al. 2004]{falcke04} Falcke, H., K{\"o}rding, E., \& Markoff, S. 2004, \aap, 414, 895
\bibitem[Farrell et al. 2009]{farrell09} {Farrell}, S.~A., {Webb}, N.~A., {Barret}, D., {Godet}, O. \& {Rodrigues}, J.~M. 2009, Natur, 460, 73
\bibitem[Fukui et al. 2012]{fukui12} Fukui, Y., Sano, H., Sato, J., et al. 2012, \apj, 746, 82
\bibitem[Gallo et al. 2005]{gallo05} Gallo E., Fender R. P., Kaiser C., et al. 2005, Natur, 436, 819
\bibitem[K{\i}z{\i}ltan et al. 2017]{kiziltan17} K{\i}z{\i}ltan, B., Baumgardt, H. \& Loeb, A. 2017, Natur, 542, 203
\bibitem[McGary et al. 2001]{mcgary01} McGary, R. S., Coil, A. L., \& Ho, P. T. P. 2001, \apj, 559, 326
\bibitem[Merloni et al. 2003]{merloni03} Merloni, A., Heinz, S., \& di Matteo, T. 2003, \mnras, 345, 1057
\bibitem[Mezcua 2017]{mezcua17} Mezcua, M. 2017, IJMPD, 26, 1730021
\bibitem[Muno et al. 2009]{muno09} Muno, M. P., Bauer, F. E., Baganoff, F. K., et al. 2009, \apjs, 181, 110
\bibitem[Nomura et al. 2018]{nomura18}  Nomura, M., Oka, T., Yamada, M., et al. 2018, \apj, 859, 29
\bibitem[Oka et al. 1998]{oka98} Oka, T., Hasegawa, T., Sato, F., Tsuboi, M., \& Miyazaki, A. 1998, \apjs, 118, 455
\bibitem[Oka et al. 2008]{oka08} Oka, T., Hasegawa, T., White, G. J., et al. 2008, \pasj, 60, 429
\bibitem[Oka et al. 2011]{oka11} Oka, T., Nagai, M., Kamegai, K., \& Tanaka, K. 2011, \apj, 732, 120
\bibitem[Oka et al. 2016]{oka16} Oka, T., Mizuno, R., Miura, K. \& Takekawa, S. 2016, \apjl, 816, L7
\bibitem[Oka et al. 2017]{oka17} Oka, T., Tsujimoto, S., Iwata, Y., Nomura, M. \& Takekawa, S. 2017, NatAs, 1, 709
\bibitem[Oka et al. 1999]{oka99} Oka, T., White, G. J., Hasegawa, T., et al. 1999, \apj, 515, 249
\bibitem[Parsons et al. 2018]{parsons18} Parsons, H., Dempsey, J. T., Thomas, H. S., et al. 2018, \apjs, 234, 22
\bibitem[Plotkin et al. 2012]{plotkin12} Plotkin R. M., Markoff S., Kelly B. C., K\"ording E., Anderson S. F. 2012, \mnras, 419, 267
\bibitem[Ravi et al. 2018]{ravi18} {Ravi}, V., {Vedantham}, H. \& {Phinney}, E.~S. 2018 \mnras, 478, L72
\bibitem[Reines et al. 2013]{reines13} Reines, A. E., Greene, J. E. \& Geha, M. 2013, \apj, 775, 116
\bibitem[Russell et al. 2007]{russell07} Russell D. M., Fender R. P., Gallo E., \& Kaiser C. R. 2007, \mnras, 376, 1341
\bibitem[Sch{\"o}del et al. 2005]{schodel05} {Sch{\"o}del}, R., {Eckart}, A., {Iserlohe}, C., {Genzel}, R. \& {Ott}, T. 2005, \apjl, 625, L111
\bibitem[Sofue 2006]{sofue06} Sofue, Y. 2006, \pasj, 58, 335
\bibitem[Soria et al. 2014]{soria14} Soria R., Long K. S., Blair W. P., et al. 2014, Science, 343, 1330
\bibitem[Strader et al. 2012]{strader12} Strader, J., Chomiuk, L., Maccarone, T. J., et al. 2012, \apjl, 750, L27
\bibitem[Takekawa et al. 2017]{takekawa17} Takekawa, S., Oka, T., Iwata, Y., Tokuyama, S. \& Nomura, M. 2017, \apjl,  843, L11
\bibitem[Tanaka 2018]{tanaka18} Tanaka, K. 2018, \apj, 859, 86
\bibitem[Tanaka et al. 2018]{tanaka_18} Tanaka, K., Nagai, M., Kamegai, K., Iino, T., \& Sakai, T. 2018, \apjs, 236, 40
\bibitem[Tanaka et al. 2015]{tanaka15} Tanaka, K., Nagai, M., Kamegai, K., \& Oka, T. 2015, \apj, 806, 130
\bibitem[Tanaka et al. 2009]{tanaka09} Tanaka, K., Oka, T., Nagai, M., \& Kamegai, K. 2009, \pasj, 61, 461
\bibitem[Tetarenko et al. 2016]{tetarenko16} Tetarenko, B. E., Bahramian, A., Arnason, R. M., et al. 2016, \apj, 825, 10
\bibitem[Tetarenko et al. 2018]{tetarenko18} Tetarenko, A. J., Freeman,P., Rosolowsky, E.W., Miller-Jones, J. C. A., \& Sivakoff, G.R. 2018, \mnras, 475, 448
\bibitem[Torii et al. 2017]{torii17} Torii, K., Hattori, Y., Hasegawa, K., et al. 2017, \apj, 835, 142
\bibitem[Tsuboi et al. 2017]{tsuboi17} Tsuboi, M., Kitamura, Y., Tsutsumi, T., et al. 2017, \apjl, 850, L5
\bibitem[Urquhart et al. 2018]{urquhart18} Urquhart, R., Soria, R., Johnston, H. M., et al. 2018, \mnras, 475, 3561 
\bibitem[Yalinewich \& Beniamini 2018]{yalinewich18} Yalinewich, A., \& Beniamini, P. 2018, \aap, 612, L9
\bibitem[Zhao et al. 2009]{zhao09} Zhao, J.-H., Morris, M. R., Goss, W. M., \& An, T. 2009, \apj, 699, 186
\bibitem[Zhao et al. 2013]{zhao13} Zhao, J.-H., Morris, M. R., \& Goss, W. M. 2013, \apj, 777, 146
\bibitem[Zhao et al. 2016]{zhao16} Zhao, J.-H., Morris, M. R., \& Goss, W. M. 2016, \apj, 817, 171
\end{thebibliography}
\end{document}